\documentstyle[12pt]{article}
\begin{document}
\begin{titlepage}
\begin{flushright}
UAB--FT--415\\
L.P.T.H.E.--ORSAY 97/10\\
April 1997
\end{flushright}
\vspace{1cm}

\begin{center}
{\bf\LARGE Pseudoscalar Conversion 
\\[0.6ex] and \\[0.6ex] Gamma-Rays from
Supernovae$^{\dagger}$
%\\[0.6ex]
}
\vspace{2cm}

\centerline{\bf 
                Eduard Mass\'o}
\vspace{0.6cm}

\centerline{Grup de F\'{\i}sica Te\`orica and IFAE,
                Universitat Aut\`onoma de Barcelona,}
\centerline{E-08193 Bellaterra, Spain}
\centerline{and}
\centerline{LPTHE, b\^at. 211, Universit\'e Paris XI, Orsay Cedex, 
France$^{\ast}$}
\vspace{3cm}

{\bf Abstract}\\
\parbox[t]{\textwidth}
{A light pseudoscalar coupled to two photons would be copiously emitted
by the core of a supernova and part of this flux would be converted to
gamma-rays by the galactic magnetic field. Measurements on the SN1987A
gamma-ray flux by the Gamma-Ray Spectrometer on the Solar Maximum
Mission satellite imply stringents bounds on such process. The improved
generation of satellite-borne detectors, like EGRET or the project
GLAST, could be able to detect a pseudoscalar-to-photon signal from a
nearby supernova.}
\end{center}
\vspace{\fill}

{\noindent\makebox[10cm]{\hrulefill}\\
\footnotesize
\makebox[1cm][r]{$^{\dagger}$}Talk at XXXIInd Rencontres de Moriond, ``Very
High Energy Phenomena in the Universe'', Les Arcs, France (1997).

\noindent
\footnotesize
\makebox[1cm][r]{$^{\ast}$}Laboratoire associ\'e au C.N.R.S. --- URA D0063.
}
\end{titlepage}
\newpage

Axions \cite{Weinberg-Wilczek} and Majorons \cite{Chikashige-Gelmini}
are examples of pseudoscalar particles arising in Particle Physics
models. Other examples are light bosons from extra-dimensional gauge 
theories \cite{Turok}, arions \cite{Anselm} and omions \cite{Sikivie}. 
Pseudoscalar particles usually couple to two photons  via the
interaction lagrangian
\begin{equation}
  {\cal L} = \frac{1}{8} \ g \phi \ 
    \varepsilon_{\mu \nu \alpha \beta}
                           F^{\mu \nu} F^{\alpha \beta} .
\end{equation}

The limits on $g$ depend on the mass $m$ of the pseudoscalar \cite{Masso}. 
For very light masses, $m \leq 10^{-9}$ eV, the best  constraints come 
from astrophysics. 

Here I consider the potential effect that in a supernova explosion one might 
detect gamma-rays before the optical fireworks. Indeed, should these
pseudoscalars exist, they are copiously produced when a supernova occurs.
For small $g$, these particles will stream away  from the supernova
core, without further interactions. In their path, the  galactic
magnetic field will convert a fraction of the flux back to gamma-ray
photons.

In order to calculate the gamma-ray flux, I will in turn consider 
$\phi$ production and the conversion effect. Finally I will
discuss the $\gamma$ detection. 

Immediately after collapse, $\phi$ is produced in the hot and superdense
core of the supernova. The Primakoff process on protons, $p \gamma
\longrightarrow p \phi$, is the main production mechanism. The number of
pseudoscalars produced per unit volume and per unit time, at temperature
$T$ and  with energy between $E_{min}$ and $E_{max}$ is
\begin{equation} \label{N(T)}
  N(T) = \int_{E_{min}}^{E_{max}} \sigma (\omega) v \ n_p 
  \ dn_{\gamma} (T,\omega) .
\end{equation}  
We integrate between $E_{min}$ and $E_{max}$ since gamma-ray detectors
are only sensitive to a fixed energy  band. In expression (\ref{N(T)})
$n_p$ and $n_{\gamma}$  are the number densities of  protons and photons
respectively, $v$ their relative  velocity and $\sigma(\omega)$ the
Primakoff cross section as a function  of the photon energy $\omega$. 

We can write the expression for the pseudoscalar flux $\Phi _{\phi} $ on
the Earth \cite{GMT}
\begin{eqnarray}
  \Phi _{\phi}  & = & 3 \times 10^{2} \; \mbox{cm}^{-2} 
           \mbox{s}^{-1} 
            \left(\frac{g}{3 \times 10^{-12}\, 
           \mbox{GeV}^{-1}}\right)^2  \\
           & & \times \left( \frac{55 \mbox{kpc}}{D} \right)^2   
           \left( \frac{R_c}{10\, \mbox{km}} \right)^3
           \left( \frac{n_p}{1.4 \times 10^{38}\, \mbox{cm}^{-3}}
           \right) 
           \left( \frac{T}{60\, \mbox{MeV}} \right)^3 
           \nonumber    \times F ,
\end{eqnarray}
where $F$ is a factor of order 1. We have normalized to the distance 
\begin{equation} \label{D}
D=55\ \mbox{kpc}
\end{equation}
since below  we will be using the SN1987A to put a numerical limit to
$g$. Also we take
\begin{eqnarray} \label{TNR}
  T & = & 60\  \mbox{MeV}  \nonumber \\
  n_p & = & 1.4 \times 10^{38}\ \mbox{cm}^{-3}  \nonumber \\          
  R_c & = & 10\  \mbox{km}
\end{eqnarray}  
as central values for the temperature, proton number density and radius,
respectively.

Let us now consider the pseudoscalar conversion. The mixing between the
photon and low mass particles in  magnetic fields leads to very
interesting phenomena. Since the pioneering work
of Sikivie \cite{Sikivie2,Sikivie}, a variety of  implications for laboratory
experiments and astrophysical observations have been investigated \cite{Masso}.

The coherent $\phi \rightarrow \gamma$ transition probability, for a 
$\phi$ beam traversing a transverse  magnetic field $B_T$ after a
distance $L$, is given by
\begin{eqnarray} \label{P}
  P\left( \phi \rightarrow \gamma \right) & = & \frac{1}{4}\
                 g^2 \ B^2_T \ L^2 \\
& = & 3 \times 10^{-4} 
           \left( \frac{g}{3 \times 10^{-12} \; \mbox{GeV} ^{-1}} \right)^2 
           \left( \frac{B_T}{1 \; \mu \mbox{G}} \right) ^2
           \left( \frac{L}{1 \; \mbox{kpc}} \right) ^2
\end{eqnarray}
The coherent effect is obtained provided
\begin{equation}
  m \leq 10^{-9} \; \mbox{eV},
\end{equation}  
so that the conversion would only be happen for such small masses.
We have adopted the model consisting in a toroidal $2\mu$G magnetic
field, leading to  $B_T = 1\, \mu$G in the SN1987A direction, and a
coherence length $L = 1$ kpc.

The photon flux $f$ (number of photons per unit area per unit time) is
then
\begin{equation}\label{F}
 f = \Phi_{\phi}P\left( \phi \rightarrow \gamma \right).
\end{equation}
We draw this flux as function of the $\gamma$ energy in the Figure. 

We finally discuss the potential detection of the gamma-ray burst. At
the time the neutrino burst from SN1987A reached the Earth, the
satellite-borne GRS was on duty measuring the incident $\gamma-$ray
flux. This measurement provides an observational limit on the
$\gamma-$ray flux coming from the supernova, and consequently on the
photons from supernova $\phi$ emission. Data from SMM \cite{Chupp} in the
energy  band $E_{min} = 25$ MeV $< E < E_{max} = 100$ MeV imply that
\begin{equation}
  \Phi_{\phi} \ P(\phi \rightarrow \gamma) \ \Delta t
  < 0.6 \; \mbox{cm}^{-2}.
\end{equation}
Using $\Delta t = 5$ s, which is roughly the characteristic signal decay
time for the neutrino burst of SN1987A, we find 
\begin{equation} \label{limit}
  g < 3 \times 10^{-12} \; \mbox{GeV}^{-1}.
\end{equation} 
This is the most stringent limit place on $g$ for small masses.
We have estimated \cite{GMT} an uncertainty of about a factor of 2 in
our limit (\ref{limit}).

The next generation of detectors, like EGRET on the Compton GRO
satellite or the project GLAST, could be able to detect a $\gamma -$ray
signal due to $\phi$ conversion from a nearby supernova collapse, for
values of $g$ allowed by our analysis.

\section*{Acknowledgments}

I acknowledge financial support from the CICYT AEN95-0815 Research Project, 
from the Theoretical Astroparticle Network under the EEC Contract No.
CHRX-CT93-0120 (Direction Generale 12 Coma), and from Direcci\'on General de 
Investigaci\'on Cientifica y Ense\~nanza Superior.

\bigskip

\begin{center}
{\Large\bf Figure Captions}
\end{center}

Differential flux $df/dE$, number of photons per cm$^2$ per 
second per MeV, as a function of photon energy $E$ in MeV. We have normalized 
to the SN1987A value $D=55$ kpc and the values displayed in (\ref{TNR})
and to $B_T=1\ \mu G$ and to $L=1$ kpc. The figure is shown for the particular
value $g=3 \times 10^{-12} \; \mbox{GeV}^{-1}$; $df/dE$ scales as $g^4$. 

\end{document}